\documentclass[groupedaddress,a4paper]{revtex4}
\usepackage{amsmath,amsthm,amssymb}
\def\today{February 13, 2002. Revised June 18, 2002.}
\newcommand{\al}{\alpha}
\newcommand{\be}{\beta}
\newcommand{\ep}{\epsilon}
\newcommand{\ga}{\gamma}

\newcommand{\la}{\lambda}

\newcommand{\vp}{\varphi}

\newcommand{\Ga}{\Gamma}
\newcommand{\La}{\Lambda}
\newcommand{\bx}{\mathbf{x}}

\newcommand{\scH}{\cH^*}
\newcommand{\szcH}{\cH^{*0}}
\newcommand{\scC}{\cC^*}
\newcommand{\scI}{\cI^*}

\newcommand{\tK}{\tilde{K}}
\newcommand{\tS}{\tilde{S}}

\newcommand{\NN}{{\mathbb N}}
\newcommand{\RR}{{\mathbb R}}
\newcommand{\ZZ}{{\mathbb Z}}
\newcommand{\cB}{{\mathcal B}}
\newcommand{\cC}{{\mathcal C}}

\newcommand{\cH}{{\mathcal H}}
\newcommand{\cI}{{\mathcal I}}
\newcommand{\cJ}{{\mathcal J}}
\newcommand{\cM}{{\mathcal M}}

\newcommand{\cR}{{\mathcal R}}
\newcommand{\cS}{{\mathcal S}}
\newcommand{\cW}{{\mathcal W}}
\newcommand{\fD}{{\mathfrak D}}
\newcommand{\fK}{{\mathfrak K}}
\newcommand{\fS}{{\mathfrak S}}
\newcommand{\fW}{{\mathfrak W}}
\newcommand{\pa}{\partial}

\newcommand{\abs}[1]{\left|#1\right|}
\def\ket#1{|#1\rangle}

\let\ds\displaystyle
\let\ni\noindent
\newcommand{\card}{\operatorname{card}}

\newcommand{\e}{{\rm e}}
\newcounter{mylc}
\renewcommand{\themylc}{\roman{mylc}}

\newtheorem{theorem}{\bf Theorem}
\newtheorem{lemma}{\bf Lemma}
\newtheorem{corollary}{\bf Corollary}
\newtheorem{proposition}{\bf Proposition}
\newcounter{rem}
\newenvironment{remark}{{\ni\em Remark~\therem.}}{\newline\stepcounter{rem}}
\begin{document}
\title{On the Sutherland Spin Model of $B_N$ Type and\\
its Associated Spin Chain}
\author{F. Finkel}
\author{D. G\'omez-Ullate}
\author{A. Gonz\'alez-L\'opez}
\author{M.A. Rodr\'{\i}guez}
\affiliation{Departamento de F\'{\i}sica Te\'orica II, Universidad
Complutense, 28040 Madrid, Spain}
\author{R. Zhdanov}
\altaffiliation{On leave of absence from Institute of Mathematics,
3 Tereshchenkivska St., 01601 Kyiv - 4 Ukraine}
\affiliation{Departamento de F\'{\i}sica Te\'orica II, Universidad
Complutense, 28040 Madrid, Spain}
\date{\today}

\begin{abstract}\hskip4mm
\begin{center}
\textbf{\normalsize Abstract}\\[3mm]
\begin{minipage}{12cm}
The $B_N$ hyperbolic Sutherland spin model is expressed in terms
of a suitable set of commuting Dunkl operators. This fact is
exploited to derive a complete family of commuting integrals of
motion of the model, thus establishing its integrability. The
Dunkl operators are shown to possess a common flag of invariant
finite-dimensional linear spaces of smooth scalar functions. This
implies that the Hamiltonian of the model preserves a
corresponding flag of smooth spin functions. The discrete spectrum
of the restriction of the Hamiltonian to this spin flag is
explicitly computed by triangularization. The integrability of the
hyperbolic Sutherland spin chain of $B_N$ type associated with the
dynamical model is proved using Polychronakos's ``freezing trick".
\end{minipage}
\end{center}
\end{abstract}
\maketitle
\section{Introduction}
\label{intro}
Since the publication of the pioneering papers of
Calogero~\cite{Ca71} and Sutherland~\cite{Su71,Su72}, the study of
solvable and integrable quantum many-body problems has become a
fruitful field of research with multiple connections in many
branches of contemporary mathematics and physics. From a
mathematical standpoint, one of the key developments in the field
was the discovery by Olshanetsky and Perelomov of an underlying
$A_N$ root system structure for both the Calogero and Sutherland
models~\cite{OP83}. The integrability of these models follows by
expressing the Hamiltonian as one of the radial parts of the
Laplace--Beltrami operator in a symmetric space associated with
the given root system. It was also shown in this paper that the
original inverse square (Calogero) and trigonometric/hyperbolic
(Sutherland) potentials arise as appropriate limits of the most
general potential in this class, given by the Weierstrass
$\wp$-function, and that integrable models associated to other
root systems also exist. The rational and trigonometric
Calogero--Sutherland (CS) models are also \emph{exactly solvable},
in the sense that their eigenfunctions and eigenvalues can be
computed algebraically. In fact, the study of the eigenfunctions
of these models has led to significant advances in the theory of
multivariate orthogonal polynomials~\cite{LV96,BF97,Du98}. Apart
from their mathematical interest, CS models have found numerous
applications in diverse areas of physics such as soliton
theory~\cite{Ka95,Po95}, fractional statistics and
anyons~\cite{BM96,CL99}, random matrix theory~\cite{TSA95}, and
Yang--Mills theories~\cite{GN94,DF98}, to name only a few.

During the last decade, CS models with internal degrees of freedom
have been actively explored by a variety of methods, including the
exchange operator formalism~\cite{MP93}, the Dunkl operators
approach~\cite{BGHP93,Ch94,Du98}, reduction by discrete
symmetries~\cite{Po99}, and construction of Lax
pairs~\cite{HW93,HW93b,IS01}. Historically, the first CS models
with spin discussed in the literature were related to the original
models of $A_N$ type introduced by Calogero and
Sutherland~\cite{HH92,BGHP93,HW93,HW93b,MP93,SS93}. The
integrability of these CS spin models was established in some
cases by relating the Hamiltonian to a quadratic combination of
Dunkl operators of $A_N$ type ~\cite{Du89,Ch91,Po92}. The $B_N$
counterpart of the $A_N$ CS spin models mentioned above were first
considered by Yamomoto~\cite{Ya95}. In this paper, the spectrum of
the rational $B_N$ spin model was explicitly determined, and its
integrability was shown by means of the Lax pair approach. In
Ref.~\cite{YT96}, Yamamoto and Tsuchiya presented an alternative
proof of the integrability of this model using Dunkl operators of
$B_N$ type. The same operators were later employed by Dunkl to
construct a complete basis of eigenfunctions~\cite{Du98}. In
contrast, the trigonometric/hyperbolic $B_N$ spin model has
received remarkably little attention. In our recent
paper~\cite{FGGRZ01b} we proved that this model is exactly
solvable in the sense of Turbiner~\cite{Tu88, Tu92}, meaning that
its Hamiltonian leaves invariant a known infinite increasing
sequence (or \emph{flag}) of finite-dimensional linear spaces of
smooth spin functions. In fact, in \cite{FGGRZ01b} we developed a
systematic method for constructing exactly (or in some cases
partially) solvable $B_N$-type CS models with spin by combining
several families of Dunkl operators. The key elements of this
method ---first introduced in the $A_N$ case in \cite{FGGRZ01}---
are: i) the definition of a new family of Dunkl operators, and ii)
the construction of a very wide class of quadratic combinations of
these operators and those in the other two families considered by
Dunkl in~\cite{Du98}.

The interest on CS spin models has been further enhanced by their
close connections with integrable spin chains of Haldane--Shastry
type~\cite{Ha88,Sh88}. Spin chains describe a fixed arrangement of
particles that interact through their spins. A well-known example
is the Heisenberg spin chain, whose spins are equally spaced and
interact only with their nearest neighbors. The Haldane--Shastry
model was actually the first one-dimensional spin chain with long
range interactions whose spectrum could be computed exactly. In
this model, the spin sites are equally spaced in a circle and
interact with each other with strength decreasing as the inverse
square of the chord distance between the sites. The integrability
of the Haldane--Shastry spin chain was proved by Fowler and
Minahan in~\cite{FM93}. Polychronakos later realized that the
commuting conserved quantities of the Haldane--Shastry spin chain
can be elegantly deduced from those of the (dynamical) Sutherland
spin model of $A_N$ type by applying what he called the ``freezing
trick''~\cite{Po93} (see also~\cite{SS93}). This corresponds to
taking the strong coupling limit in the Sutherland spin model and
restricting to states with no momentum excitations, so that the
internal degrees of freedom remain the only relevant variables in
the problem and the particles are ``frozen'' at their classical
equilibrium positions. This observation is, in principle, valid
for any integrable spin Calogero--Sutherland model. For instance,
in Ref.~\cite{Po93} the freezing trick is applied to the spin
Calogero model with rational interaction to construct a new
integrable spin chain of rational type in which the sites are no
longer equally spaced. The spectrum of this chain was later
calculated by Frahm~\cite{Fr93} and Polychronakos~\cite{Po94}.
Bernard, Pasquier and Serban~\cite{BPS95} studied the spin chain
associated with the trigonometric Sutherland model, establishing
its integrability for certain values of the parameters in the
Hamiltonian.

The aim of this paper is twofold. In the first place, we prove the
integrability of the hyperbolic Sutherland spin model of $B_N$
type, from which we are also able to deduce the integrability of
the spin chain associated with this model. Secondly, we give an
explicit formula for the eigenvalues of the dynamical model whose
corresponding (square-integrable) eigenfunctions lie in the
invariant flag mentioned above. The paper is organized as follows.
In Section \ref{integrability} we introduce a family of commuting
Dunkl operators of $B_N$ type. We show that the sums of even
powers of these Dunkl operators generate a complete set of
commuting integrals of motion of the Hamiltonian of the model. The
commutation relations satisfied by the Dunkl operators and the
usual permutation and sign reversing operators, which possess a
richer structure than in the rational case, play a key role in the
proof of this result. In Section~\ref{spectrum} we analyze the
spectrum of the Hamiltonian for any value of the spin. Our
analysis is based on the fact that the Dunkl operators leave
invariant a flag of finite-dimensional linear spaces of smooth
scalar functions. This flag yields a corresponding invariant flag
of smooth spin functions for the hyperbolic $B_N$ Sutherland
Hamiltonian. We construct a partially ordered basis of this
``spin" flag in which the Hamiltonian is represented by a
triangular matrix. In this way we can explicitly compute the
eigenvalues of the restriction of the Hamiltonian to the
(finite-dimensional) intersection of the spin flag with the
Hilbert space of the system. We shall use the term
\emph{algebraic} in what follows to refer to the these eigenvalues
and its corresponding eigenfunctions. It remains an open problem
to determine whether the algebraic sector of the spectrum actually
coincides with the discrete spectrum. We also study in detail the
(algebraic) ground state, determining its degeneracy for all
values of the spin. In Section \ref{chain} we define the spin
chain associated with the hyperbolic Sutherland spin model of
$B_N$ type, and apply the freezing trick to derive a complete
family of commuting integrals of motion of this chain.

\section{Integrability of the Sutherland Spin Model of $B_N$ Type}
\label{integrability}
The Hamiltonian of the {\em hyperbolic $B_N$ Sutherland spin
model} is defined by
\begin{equation}
\label{HBNSuth}
\begin{aligned}
H^*=-\sum_i \pa_{x_i}^2 &+ 2a\,\sum_{i<j}\left[\sinh^{-2}
x_{ij}^-\,(a+S_{ij})+
\sinh^{-2} x_{ij}^+\,(a+\tS_{ij})\right]\\
&{}+b\,\sum_i \sinh^{-2}\!x_i\,(b+S_i) -b'\,\sum_i
\cosh^{-2}\!x_i\,\big(b'+S_i\big)\,,
\end{aligned}
\end{equation}
where $x_{ij}^\pm=x_i\pm x_j$ and $a,b,b'$ are real parameters.
Here and in what follows, any summation or product index without
an explicit range will be understood to run from $1$ to $N$,
unless otherwise constrained. The operators $S_{ij}$ and $S_i$ in
Eq.~\eqref{HBNSuth} act on the finite-dimensional Hilbert space
\begin{equation}
\label{spinbasis} \cS=\Big\langle\,|s_1,\dots,s_N\rangle\;\Big|\;
s_i=-M,-M+1,\dots,M;\; M\in\frac12\NN\,\Big\rangle,
\end{equation}
associated to the particles' internal degrees of freedom, as
follows:
\begin{equation}
\begin{aligned}
& S_{ij}|s_1,\dots,s_i,\dots,s_j,\dots,s_N\rangle=|s_1,\dots,
s_j,\dots,s_i,\dots,s_N\rangle\,,\\
&
S_i|s_1,\dots,s_i,\dots,s_N\rangle=|s_1,\dots,-s_i,\dots,s_N\rangle\,.
\label{SS}
\end{aligned}
\end{equation}
We have also used the customary notation $\tS_{ij} =S_iS_jS_{ij}$.

The operators $S_{ij}$ and $S_i$ are represented in $\cS$ by
$(2M+1)^N$-dimensional Hermitian matrices, and obey the following
algebraic relations:
\begin{equation}
\begin{gathered}
    S_{ij}^2=1,\qquad S_{ij}S_{jk}=S_{ik}S_{ij}=S_{jk}S_{ik},\qquad
    S_{ij}S_{kl}=S_{kl}S_{ij},\\
    S_i^2=1,\qquad S_iS_j=S_jS_i\,,\qquad S_{ij}S_k=S_k S_{ij},\qquad
    S_{ij}S_j=S_i S_{ij}\,,
\end{gathered}
    \label{SSalg}
\end{equation}
where the indices $i,j,k,l$ take distinct values in the range
$1,\dots,N$. The algebra $\fS$ generated by the operators $S_{ij},
S_i$ is thus isomorphic to the group algebra of the Weyl group
$\cW_N$ of type $B_N$, also known as the hyperoctahedral group.

We shall also make use of the permutation operators
$K_{ij}=K_{ji}$ and the sign reversing operators $K_i$ ($1\le i\ne
j\le N$), whose action on a function $f(\bx)$, with
$\bx=(x_1,\dots,x_N)\in\RR^N$, is defined as follows:
\begin{equation}
\begin{aligned}
& (K_{ij}f)(x_1,\dots,x_i,\dots,x_j,\dots,x_N)=f(x_1,\dots,
x_j,\dots,x_i,\dots,x_N)\,,\\
& (K_i f)(x_1,\dots,x_i,\dots,x_N)=f(x_1,\dots,-x_i,\dots,x_N)\,.
\label{KR}
\end{aligned}
\end{equation}
The operators $K_{ij}$ and $K_i$ obey algebraic identities
analogous to \eqref{SSalg}. We shall denote by $\fK\simeq\fS$ the
algebra generated by the coordinate permutation and sign changing
operators $K_{ij}$ and $K_i$. Note also that the operators
$\Pi_{ij}=K_{ij}S_{ij}$ and $\Pi_i=K_i S_i$ generate an algebra
$\fW$ isomorphic to $\fK$ and $\fS$. From now on we shall identify
the abstract group $\cW_N$ with its realizations generated by the
operators $K_{ij},K_i$ on $C^\infty(\RR^N)$, $S_{ij},S_i$ on
$\cS$, or $\Pi_{ij},\Pi_i$ on $C^\infty(\RR^N)\otimes\cS$,
depending on the context.

The Hamiltonian \eqref{HBNSuth} describes a system of $N$
identical particles, whose physical states are therefore either
totally symmetric or totally antisymmetric under particle
exchange. Moreover, since $H^*$ clearly commutes with the family
of commuting operators $\Pi_i$ ($i=1,\dots,N$), we can choose a
basis of common eigenfunctions of $H^*$ and all the operators
$\Pi_i$. Given an element $\psi_k$ of this basis, it follows from
the commutation relations of the sign reversing operators $\Pi_i$
with the permutation operators $\Pi_{ij}$ that $\Pi_i\psi_k =
\ep_k\psi_k$, independently of $i$. In principle, the parity
$\ep_k$ could depend on $k$. However, we shall see in the
following section that all the algebraic eigenfunctions have the
same parity, and that this parity is determined by the sign of the
parameter $b$. From now on we shall assume, for definiteness, that
we are dealing with a system of fermions whose algebraic states
are also antisymmetric under sign reversal of each particle's
spatial and internal coordinates. This covers what is perhaps the
physically most interesting case, namely that of a system of spin
$1/2$ particles, for which the internal degrees of freedom are
naturally interpreted as the particles' spin. The results of this
paper can be easily modified to treat any other choice of the
particles' statistics and parity.

In the rest of this section we shall prove the integrability of
the model \eqref{HBNSuth} by expressing the Hamiltonian in terms
of the following family of $B_N$-type Dunkl operators:
\begin{multline}\label{Jexp}
J_i=\pa_{x_i} -a\sum_{j\ne i}\Big[(1+\coth x_{ij}^-)\,K_{ij}
+(1+\coth x_{ij}^+)\,\tK_{ij}\Big]\\
-\big[b\,(1+\coth x_i)+b'\,(1+\tanh
x_i)\big]K_i+2a\sum_{j<i}K_{ij}\,,
\end{multline}
where $\tK_{ij}=K_iK_jK_{ij}$. The operators $J_i$ are related to
the operators introduced by Yamamoto \cite{Ya95} in connection
with the trigonometric $BC_N$ spin Sutherland model.

A key property of the Dunkl operators \eqref{Jexp} is their
commutativity:
\begin{equation}
\label{Jcomm} [J_i,J_j]=0\,,\qquad i,j=1,\dots,N\,.
\end{equation}
We shall also make use of the following commutation relations
between the operators $K_{ij},K_i$ and the operators $J_i$:
\begin{align}
\label{KijJk} &[K_{ij}, J_k] = \begin{cases}
2a(K_{jk}-K_{ik})K_{ij}\,,\quad& i<k<j\\[1mm]
0\,,& \text{otherwise}
\end{cases}\\[2mm]
&K_{ij}J_i-J_jK_{ij} = -2a\Big(1+\sum_{i<l<j}K_{ij}K_{il}\Big),\\[2mm]
&K_{ij}J_j-J_iK_{ij} = 2a\Big(1+\sum_{i<l<j}K_{ij}K_{jl}\Big),\\[2mm]
&[K_i,J_j]=2aK_{ij}(K_j-K_i)\,,\qquad
[K_j,J_i]=0\,,\\[2mm]
\label{KiJi} &\{K_i,J_i\}=-2(b+b')-2a\sum_{l>i}K_{il}(K_i+K_l)\,,
\end{align}
where $i<j$ and $k$ are three distinct indices ranging from $1$ to
$N$. Note that the Dunkl operators of rational type considered in
Ref.~\cite{YT96} satisfy the latter equations with the right-hand
side replaced by zero. The nonvanishing of the right-hand side of
Eqs.~\eqref{KijJk}--\eqref{KiJi} gives rise to some nontrivial
technical points in the proof of the integrability of the
Hamiltonian \eqref{HBNSuth}, as we shall see below.

Another important property of the Dunkl operators $J_i$ is that
they preserve the linear space
\begin{equation}
\label{Rdef} \cR_m = \Big\langle\mu(\bx)\exp\Big(2\sum_i n_i
x_i\Big)\;\Big|\; n_i=-m,-m+1,\dots, m\,,\quad i=1,\dots,N
\Big\rangle\,,
\end{equation}
where
\begin{equation}\label{mu}
\mu(\bx) = \prod_{i<j}\big|\sinh x_{ij}^-\, \sinh x_{ij}^+
\big|^a\cdot \prod_i\big|\sinh x_i\big|^b \big|\cosh
x_i\big|^{b'}\,,
\end{equation}
for any nonnegative integer $m$. This fact will prove crucial for
the calculation of the spectrum of the hyperbolic $B_N$ Sutherland
spin model.

We define a mapping ${}^*:\fD\otimes\fK\to\fD\otimes\fS$, where
$\fD$ denotes the algebra of linear differential operators on
$C^\infty(\RR^N)$, as follows. If $D\in\fD$, we set
\begin{equation}
\label{star} \big(D\,K_{\al_1}\cdots
K_{\al_r}\big)^*=(-1)^rD\,S_{\al_r}\cdots S_{\al_1}\,,
\end{equation}
where $\al_k$ stands for $ij$ or $i$. The mapping $^*$ is then
linearly extended to $\fD\otimes\fK$. For instance, the
Hamiltonian of the hyperbolic $B_N$ Sutherland spin model
\eqref{HBNSuth} is obtained by applying the ``star" mapping to the
operator
\begin{equation}
\label{YamK}
\begin{aligned}
H=-\sum_i \pa_{x_i}^2 &+ 2a\,\sum_{i<j}\left[\sinh^{-2}
x_{ij}^-\,(a-K_{ij})+
\sinh^{-2} x_{ij}^+\,(a-\tK_{ij})\right]\\
&{}+b\,\sum_i \sinh^{-2}\!x_i\,(b-K_i) -b'\,\sum_i
\cosh^{-2}\!x_i\,\big(b'-K_i\big)\,.
\end{aligned}
\end{equation}

Let $\La_0$ be the antisymmetrisation operator, defined by the
relations $\La_0^2=\La_0$ and $\Pi_{ij}\La_0=-\La_0$,
$j>i=1,\ldots,N$. More explicitly,
$$
\La_0=\frac1{N!}\sum_{l=1}^{N!}\ep_l\,P_l\,,
$$
where $P_l$ denotes an element of the realization of the symmetric
group generated by the operators $\Pi_{ij}$, and $\ep_l$ is the
signature of $P_l$. For instance, if $N=2,3$ the antisymmetriser
$\La_0$ is given by
\begin{align*}
& N=2:\qquad \La_0=\frac12\,\big(1-\Pi_{12}\big)\,,\\[1mm]
& N=3:\qquad
\La_0=\frac16\,\big(1-\Pi_{12}-\Pi_{13}-\Pi_{23}+\Pi_{12}\Pi_{13}+
\Pi_{12}\Pi_{23}\big)\,.
\end{align*}
The total antisymmetriser $\La$ with respect to the action of the
operators $\Pi_{ij}$ and $\Pi_i$ is determined by the relations
$\La^2=\La$ and
\begin{equation}
\label{Ladef}
    \Pi_{ij}\La=-\La,\qquad \Pi_i\La=-\La,\qquad
    j>i=1,\ldots,N.
\end{equation}
It may be easily shown that
$$
\La=\frac1{2^N}\bigg(\prod_i (1-\Pi_i)\bigg)\La_0\,.
$$

Since $K_{ij}^2=K_i^2=1$, the relations \eqref{Ladef} are
equivalent to
\begin{equation}
K_{ij}\La=-S_{ij}\La,\qquad K_i\La=-S_i\La,\qquad j>i=1,\ldots,N.
    \label{La}
\end{equation}
>From these relations and the definition of the star mapping it
follows immediately that
\begin{equation}
\label{ALambda} A\,\La = A^*\La
\end{equation}
for every operator $A\in\fD\otimes\fK$. The proof of the
integrability of the hyperbolic $B_N$ Sutherland spin model is
based on the following lemmas.
\begin{lemma}
\label{lemma.AL0} If $B\in\fD\otimes\fS$ satisfies $B \La=0$, then
$B=0$.
\end{lemma}
\begin{proof}
The operator $B\in\fD\otimes\fS$ is of the form
\begin{equation}
B = \sum_{i\in I} f_i(\bx)\,B_i\,\pa^i\,,
\end{equation}
where $B_i\in\fS$, $f_i\in C^\infty(\RR^N)$, $i=(i_1,\dots,i_N)$
is a multiindex belonging to a finite subset $I\subset\NN_0^N$
(with $\NN_0=\{0,1,2,\dots\}$), and $\pa^i =
\pa_1^{i_1}\cdots\pa_N^{i_N}$. Let us denote by $W_l$, with $l\in
L=\{1,\dots,2^N N!\}$, the elements of the realization of the Weyl
group $\cW_N$ generated by the operators $\Pi_{ij}$ and $\Pi_i$.
The action of the total antisymmetrisation operator $\La$ over a
factored state $\psi=\vp(\bx)\,\ket s$, with $\vp\in
C^\infty(\RR^N)$ and $\ket s\in\cS$, is given by
\begin{equation}
\La\,\psi = \sum_{l\in L} \ep_l\,(W_l\vp)\,W_l\ket{s}\,,
\end{equation}
where $\ep_l=\pm1$ is the parity of the total number of generators
$\Pi_{ij}$, $\Pi_i$ in any decomposition of $W_l$. By hypothesis,
\begin{equation}
\label{ALa} B(\La\,\psi) = \sum_{i\in I,l\in L}
\ep_l\,f_i(\bx)\pa^i\big(W_l\vp\big)\cdot B_i(W_l\ket{s}) = 0\,.
\end{equation}
Applying the latter equation to a family of functions
$\big\{\vp_j(\bx)\big\}_{j\in I\times L}$ satisfying the condition
\begin{equation}
\det\Big[\pa^i\big(W_l\vp_j\big)\Big]^{i\in I,l\in L}_{j\in
I\times L}\ne0\,,
\end{equation}
we obtain
$$
B_i(W_l\ket{s})=0\,,\qquad\text{for all }\: i\in I,\:l\in L\,.
$$
In particular (taking $l\in L$ so that $W_l$ is the identity)
$B_i\ket s=0$ for all $i\in I$. Since $\ket s\in\cS$ is arbitrary,
it follows that $B_i=0$ for all $i\in I$, and hence $B$ vanishes
identically.
\end{proof}
\begin{lemma}
\label{lem.fund} If $B\in\fD\otimes\fK$ commutes with $\La$ then
$(AB)^* = A^*B^*$ for all $A\in\fD\otimes\fK$.
\end{lemma}
\begin{proof}
Using Eq.~\eqref{ALambda} and the hypothesis repeatedly we obtain:
$$
(AB)^*\La = AB\La = A\La B = A^*\La B = A^* B \La = A^* B^* \La\,.
$$
The statement follows from the previous lemma.
\end{proof}
We shall often make use of the following immediate consequence of
Lemma \ref{lem.fund}:
\begin{lemma}
\label{lem.comm} If $A,B\in\fD\otimes\fK$ commute with $\La$ then
$[A,B]^* = [A^*,B^*]$.
\end{lemma}

We shall now construct a complete family of commuting integrals of
motion for the hyperbolic $B_N$ Sutherland spin Hamiltonian
\eqref{HBNSuth}. The construction is based on the observation that
the operator $H$ in Eq.~\eqref{YamK} can be expressed as
\begin{equation}\label{H}
H=-\sum_i J_i^2\,.
\end{equation}
By \eqref{Jcomm}, the operators
\begin{equation}\label{Ip}
  I_p=\sum_i J_i^{2p}\,,\qquad p\in\NN\,,
\end{equation}
commute with one another. In view of the previous lemma, it
suffices to prove that $I_p$ commutes with the total
antisymmetriser $\La$ for all $p\in\NN$ to conclude that the star
operators $I_p^*$ ($p\in\NN$) form a commuting family of integrals
of motion of $H^*=-I_1^*$. We shall in fact prove the following
stronger result:
\begin{lemma}
\label{lemma.IpK} The operators $I_p$ $(p\in\NN)$ commute with the
permutation and sign changing operators $K_{ij}$ and $K_i$.
\end{lemma}
\begin{proof}
First of all, the elementary permutation $K_{i,i+1}$ commutes with
$I_p$ for $i=1,\dots,N-1$. Indeed, $K_{i,i+1}$ commutes with $J_j$
for $j\ne i,i+1$ by \eqref{KijJk}, while for $j=i,i+1$ we have
\begin{align*}
K_{i,i+1}J_i^{2p}&=J^{2p}_{i+1} K_{i,i+1}
-2a\sum_{r=0}^{2p-1} J_i^{2p-r-1} J^{r}_{i+1},\\
K_{i,i+1} J^{2p}_{i+1} &=J_{i}^{2p} K_{i,i+1} +2a\sum_{r=0}^{2p-1}
J_i^{2p-r-1} J^{r}_{i+1}.
\end{align*}
Since an arbitrary permutation can be expressed as the product of
elementary permutations, this shows that $K_{ij}$ commutes with
$I_p$ for all $i\ne j$. Secondly, the sign reversing operator
$K_N$ commutes with $I_p$ for all $p$, since
\begin{align*}
& K_N J_i= J_i K_N,\quad \text{if}\quad i<N,\\
\intertext{while} & K_N J^2_N=- J_NK_N J_N-2(b+b') J_N=
J^2_NK_N\,.
\end{align*}
This implies that $K_i$ commutes with $I_p$ for an arbitrary
$i=1,\dots,N$, since
$$
0=K_{iN}\big[K_NK_{iN},I_p\big]=K_{iN}\big[K_{iN}K_i,I_p\big]=\big[K_i,I_p\big].
$$
\end{proof}
The operators $I_p^*$ ($p\in\NN$) thus form an infinite commuting
family. Moreover, by examining the terms of highest order in the
partial derivatives one can easily conclude that the set
$\{I_p^*\}_{p=1}^N$ is algebraically independent. We have thus
proved the main result of this section:
\begin{theorem}
\label{thm.1} The operators $\{I_p^*\}_{p=1}^N$ form a complete
family of commuting integrals of motion for the hyperbolic $B_N$
Sutherland spin Hamiltonian $H^*=-I_1^*$.
\end{theorem}
We also note that the constants of motion $I_p^*$ ($p\in\NN$)
commute with the total permutation and sign changing operators
$\Pi_{ij}$ and $\Pi_i$. This is a consequence of Lemma
\ref{lemma.IpK} and the following general fact:
\begin{lemma}
\label{IpstarK} If $A\in\fD\otimes\fK$ commutes with $K_{ij}$
(resp.~$K_i$) then $A^*$ commutes with $\Pi_{ij}$ (resp.~$\Pi_i$).
\end{lemma}
\begin{proof}
We can write
\begin{equation}
\label{Aexp} A = \sum_{\ga\in \Ga} D_\ga K_\ga\,,
\end{equation}
where $K_\ga$ is a monomial in $K_{kl}$ and $K_l$, $D_\ga\in\fD$,
and $\Ga$ is a finite set such that $\{K_\ga \mid \ga\in\Ga\}$ is
linearly independent. By hypothesis
\begin{equation}
\label{KAKexp} A = K_{ij} A K_{ij} = \sum_{\ga\in\Ga}
K_{ij}(D_\ga)\,K_{ij}K_\ga K_{ij}\,
\end{equation}
where $K_{ij}(D_\ga)$ is the image of $D_\ga$ under the natural
action of $K_{ij}$ in $\fD$. Comparing \eqref{KAKexp} with
\eqref{Aexp} we conclude that for each $\ga\in\Ga$ there exists
$\ga'\in\Ga$ such that $K_{ij}K_\ga K_{ij}= K_{\ga'}$, and
$$
K_{ij}(D_\ga)=D_{\ga'}\,.
$$
On the other hand we have
$$
\Pi_{ij}A^*\Pi_{ij} = \sum_{\ga\in\Ga}
K_{ij}(D_\ga)\,S_{ij}K_\ga^* S_{ij} = \sum_{\ga\in\Ga}
K_{ij}(D_\ga)\,K_{\ga'}^* = \sum_{\ga\in\Ga} D_{\ga'}\,K_{\ga'}^*
\,.
$$
Since $(\ga')'=\ga$, we have $\Ga'=\Ga$, and therefore the
right-hand side of the previous formula equals $A^*$. The equality
$\Pi_i A^* \Pi_i = A^*$ is established in a similar way.
\end{proof}

\section{Spectrum of the Sutherland Spin Model of $B_N$ Type}
\label{spectrum}
We shall now study the algebraic sector of the spectrum of the
$B_N$-type Sutherland spin model \eqref{HBNSuth}. The starting
point in our discussion is the invariance under $H$ of the space
$\cR_m$ for all $m=0,1,\dots$, which is an immediate consequence
of Eq.~\eqref{H} and the definition of the operators $J_i$. We
shall construct a basis of the $H$-invariant space $\cR_m$ with
respect to which the matrix of $H|_{\cR_m}$ is upper triangular,
thereby obtaining an exact formula for the spectrum of this
operator.

To derive the spectrum of $H^*$ from that of $H$, we shall make
use of the identity
\begin{equation}
\label{HstH} H^*\big[\La\big(\vp\ket s\big)\big] =
\La\big[(H\vp)\ket s\big]\,,
\end{equation}
where $\vp\in C^\infty(\RR^N)$ and $\ket s\in\cS$. The latter
identity, which is an immediate consequence of Eq.~\eqref{ALambda}
and Lemma~\ref{lemma.IpK}, implies that the spaces
\begin{equation}
\cM_m = \La(\cR_m\otimes\cS)\,,\qquad m=0,1,\dots,
\end{equation}
are invariant under $H^*$. From the basis of $\cR_m$
triangularizing $H|_{\cR_m}$ we shall construct a basis of $\cM_m$
with respect to which $H^*|_{\cM_m}$ is also represented by an
upper triangular matrix. In this way we shall determine the
spectrum of the Sutherland spin model of $B_N$ type
\eqref{HBNSuth}.

Let us start by computing the spectrum of the operator $H$.
Following closely the approach of Ref.~\cite{BGHP93} for spin
models of $A_N$ type, we shall define a suitable partial ordering
in the set of (scaled) exponential monomials
\begin{equation}
\label{fn} f_n(\bx) = \mu(\bx)\exp\Big(2\sum_i n_i
x_i\Big)\,,\qquad n=(n_1,\dots,n_N)\,,\quad -m\le n_i\le m\,,
\end{equation}
spanning the subspaces $\cR_m$. We shall then show that the
operator $H$ is represented by a triangular matrix in any
partially ordered basis of $\cR_m$.

The partial ordering in the basis \eqref{fn} is defined as
follows. Given a multiindex $n=(n_1,\dots,n_N)\in\ZZ^N$, we define
the nonnegative and nonincreasing multiindex $[n]$ by
\begin{equation}
\label{brn} [n]=\big(\abs{n_{i_1}},\dots,\abs{n_{i_N}}\big)\,,
\qquad\text{where}\quad \abs{n_{i_1}}\ge\dots\ge\abs{n_{i_N}}\,.
\end{equation}
If $n,n'\in\big[\ZZ^N\big]$ are nonnegative and nonincreasing
multiindices, we shall say that $n\prec n'$ if
$n_1-n_1'=\dots=n_{i-1}-n_{i-1}'=0$ and $n_i <n_i'$. For two
arbitrary multiindices $n,n'\in\ZZ^N$, by definition $n\prec n'$
if and only if $[n]\prec[n']$. The partial ordering $\prec$ in
$\ZZ^N$ induces a partial ordering in the exponential monomial
basis \eqref{fn}, namely $f_n\prec f_{n'}$ if and only if $n\prec
n'$. The action of the Weyl group on the basis \eqref{fn}
preserves this partial ordering, i.e., if $f_n\prec f_{n'}$ then
$Wf_n\prec Wf_{n'}$ for all $W\in\cW_N$.

If $n=(n_1,\dots,n_N)\in\big[\ZZ^N\big]$ and $s\in\ZZ$, we shall use the
following notation:
\begin{align*}
\#(s) &= \card\{i:n_i=s\}\,,\\
\ell(s) &= \min\{i:n_i=s\}\,,
\end{align*}
with $\ell(s)=+\infty$ if $n_i\ne s$ for all $i=1,\dots,N$.
For instance, if $n=(5,2,2,1,1,1,0)$ then $\#(1)=3$ and
$\ell(1)=4$. The computation of the spectrum of $H$ is based on
the following result:
\begin{proposition}
\label{prop.Jifn} If $n\in\big[\ZZ^N\big]$ is a nonnegative and
nonincreasing multiindex, the following identity holds:
\begin{equation}
\label{Jifnstruct} J_i f_n = \la_{n,i} f_n + \sum_{\substack{n'\in
\ZZ^N\\ n'\prec \,n}} c_{n,i}^{n'}\,f_{n'}\,,
\end{equation}
where $c_{n,i}^{n'}\in\RR$ and
\begin{equation}\label{la}
\la_{n,i} =
\begin{cases}
2n_i+b+b'+2a\big(N+i+1-\#(n_i)-2\ell(n_i)\big),\quad& n_i>0\\[3pt]
-b-b'+2a(i-N)\,,\quad& n_i=0.
\end{cases}
\end{equation}
\end{proposition}
\begin{proof}
After some algebra one readily obtains the following expression:
\begin{align}
J_i f_n = f_n\Bigg[&2n_i+b+b'+2a(N-1)-2a\sum_{j<i}\bigg(
\frac{\al_{ij}^{n_j-n_i}-1}{\al_{ij}-1}+
\frac{\be_{ij}^{1-n_i-n_j}-1}{\be_{ij}-1}\bigg)\notag\\
\label{Jifn} &{}-2a\sum_{j>i}\bigg(
\frac{\al_{ij}^{1+n_j-n_i}-1}{\al_{ij}-1}+
\frac{\be_{ij}^{1-n_i-n_j}-1}{\be_{ij}-1}\bigg)\\
&-2b\,\frac{z_i^{1-2n_i}-1}{z_i-1}
-2b'\,\frac{z_i^{1-2n_i}+1}{z_i+1}\Bigg]\,,\notag
\end{align}
where
$$
z_i=\e^{2x_i}\,,\qquad \al_{ij}=z_i z_j^{-1}\,,\qquad \be_{ij}=z_i
z_j\,.
$$
Consider, for instance, the first term in $\al_{ij}$ in
Eq.~\eqref{Jifn}. Since $j<i$, $n_j\geq n_i$. If $n_j=n_i$ this
term vanishes. If $n_j>n_i$ we have
\begin{equation}\label{fnterms}
f_n\,\frac{\al_{ij}^{n_j-n_i}-1}{\al_{ij}-1}
=f_n\bigg(1+\sum_{r=1}^{n_j-n_i-1}z_j^{-r} z_i^{r}\bigg)\,,
\end{equation}
where the last sum only appears if $n_j-n_i>1$. In this case we
have $0<\max\{n_j-r,n_i+r\}<n_j$ for all $r=1,\dots,n_j-n_i-1$, so
the multiindices of the monomials in the summation symbol in
Eq.~\eqref{fnterms} satisfy
$$
(n_1,\dots,n_j-r,\dots,n_i+r,\dots,n_N)\prec n\,.
$$
It may be likewise verified that the multiindices $n'$ of the
monomials arising from the remaining terms in Eq.~\eqref{Jifn}
either coincide with $n$ or satisfy $n'\prec n$. The value of
$\la_{n,i}$ given in Eq.~\eqref{la} can be computed by evaluating
the constant part of the expression in square brackets in the
right-hand side of Eq.~\eqref{Jifn}. For instance, the first term
in $\al_{ij}$ in Eq.~\eqref{Jifn} contributes the quantity
$-2a\big(\ell(n_i)-1\big)$ to $\la_{n,i}$.
\end{proof}
Note that Eq.~\eqref{Jifnstruct} does not hold if $n$ does not
belong to $\big[\ZZ^N\big]$, so that Proposition~\ref{prop.Jifn}
in general does not determine the spectrum of the restriction of
$J_i$ to $\cR_m$. On the other hand, for an arbitrary multiindex
$n\in\ZZ^N$ we shall only need the following weaker result:
\begin{corollary}
\label{cor.Jifn} If $n\in\ZZ^N$ then
\begin{equation}
\label{Jifnarb} J_if_n = \sum_{\substack{n'\in\ZZ^N\\ [n']\preceq
[n]}}\ga_{n,i}^{n'}\,f_{n'}\,,
\end{equation}
for some real constants $\ga_{n,i}^{n'}$.
\end{corollary}
\begin{proof}
We have
$$
J_if_n = J_i W f_{[n]}\,,
$$
where $W$ is any element of $\cW_N$ such that $f_n = Wf_{[n]}$.
Eq.~\eqref{Jifnarb} then follows from the previous proposition,
the commutation relations \eqref{KijJk}--\eqref{KiJi} and the
invariance of the partial ordering ${}\prec{}$ under the action of
the Weyl group.
\end{proof}
The algebraic spectrum of $H$ can be computed in closed form using
the previous results, which imply the following proposition:
\begin{proposition}
For all $n\in\ZZ^N$ the following identity holds:
\begin{equation}
\label{Hgfn} H f_n = -\sum_i\la_{[n],i}^2 f_n +
\sum_{\substack{n'\in \ZZ^N\\ n'\prec
\,n}}c_n^{n'}\,f_{n'}\,,\qquad \text{with }c_n^{n'}\in\RR\,.
\end{equation}
\end{proposition}
\begin{proof}
Let $W$ be any element of $\cW_N$ such that $f_n = W f_{[n]}$.
Since $H=-I_1$ commutes with $W$ by Lemma \ref{lemma.IpK}, from
\eqref{Jifnstruct} we obtain
$$
H f_n = W H f_{[n]} = -\sum_i\la_{[n],i}^2 f_n
-\sum_{\substack{n'\in \ZZ^N\!,\;i\\
n'\prec\,[n]}}\la_{[n],i}c_{[n],i}^{n'}\,Wf_{n'}
-\sum_{\substack{n'\in \ZZ^N\!,\;i\\ n'\prec\,[n]}}
c_{[n],i}^{n'}\,WJ_if_{n'}\,.
$$
Eq.~\eqref{Hgfn} follows immediately from the latter equation,
Corollary \ref{cor.Jifn} and the invariance of the partial
ordering ${}\prec{}$ under the action of the Weyl group.
\end{proof}
Let $\cB_m=\big\{f_{n(j)}\:|\:j=1,\dots, (2m+1)^N\big\}$ be any
exponential monomial basis of the linear space $\cR_m$ partially
ordered according to $\prec$, i.e., such that if $n(j)\prec n(k)$
then $j<k$. The previous proposition implies that the matrix of
the restriction of $H$ to $\cR_m$ with respect to $\cB_m$ is upper
triangular. The eigenvalues of this matrix are its diagonal
elements
\begin{equation}
\label{energy} E_n=-\sum_i\la_{[n],i}^2\,; \qquad -m\le n_j\le
m\,,\quad j=1,\dots,N\,.
\end{equation}

It should be noted, however, that the algebraic eigenfunctions of
$H$ must satisfy appropriate boundary conditions that we shall now
discuss. In the first place, since the potential of the $B_N$-type
spin Sutherland Hamiltonian \eqref{HBNSuth} diverges on the
hyperplanes $x_i\pm x_j=0$, $1\le i\le j\le N$, as $(x_i\pm
x_j)^{-2}$, we must require that the eigenfunctions of $H$ vanish
faster than $(x_i\pm x_j)^{1/2}$ near these hyperplanes. This
yields the conditions (cf.~Eq.~\eqref{mu})
\begin{equation}
\label{abcond} a,b>\frac12\,.
\end{equation}
Secondly, the eigenfunctions must be square-integrable on their
domain, which (without loss of generality) shall be taken as the
open set $X\subset\RR^N$ given by
\begin{equation}
0<x_N<\dots<x_1\,.
\end{equation}
The algebraic eigenfunctions lying in $\cR_m$ will satisfy this
condition if and only if
\begin{equation}
\label{mcond} \frac12(b+b')+a(N-1)+m<0\,.
\end{equation}
The latter inequality implies that the number of algebraic levels
of $H$ is finite, since $m$ cannot exceed the integer $m_1$
defined by
\begin{equation}
\label{m1} m_1 =
\max\Big\{m\in\NN_0\;\Big|\;\frac12(b+b')+a(N-1)+m<0\Big\}\,.
\end{equation}
Note, in particular, that there are no algebraic eigenfunctions
unless the parameters in the potential verify the inequality
\begin{equation}
\label{abcond2} \frac12(b+b')+a(N-1)<0\,.
\end{equation}
>From now on, we shall work on the maximal $H$-invariant subspace
$\cR_{m_1}$.
\begin{remark}
We could also have considered algebraic eigenfunctions of the
Hamiltonian \eqref{HBNSuth} antisymmetric under permutations but
even under sign reversals. On these eigenfunctions, the action of
the operators $S_{ij}$ and $S_i$ coincides with that of the
operators $-K_{ij}$ and $K_i$, respectively. Therefore
Eq.~\eqref{star} in the definition of the star mapping should be
replaced by
$$
\big(D\,K_{\al_1}\cdots
K_{\al_r}\big)^*=(-1)^{r'}D\,S_{\al_r}\cdots S_{\al_1}\,,
$$
where $r'$ is the number of permutation operators in the monomial
$K_{\al_1}\dots K_{\al_r}$. As a consequence, the Hamiltonian
\eqref{HBNSuth} is the image under the new star mapping of the
operator $H(-b,-b')$, with $H(b,b')$ given by Eq.~\eqref{YamK}. It
follows from Eqs.~\eqref{abcond} and \eqref{abcond2} that $H^*$
possesses algebraic eigenfunctions of even parity if and only if
\begin{equation}
\label{abcond3} a>\frac12\,,\qquad b<-\frac12\,,\qquad
-\frac12(b+b')+a(N-1)<0\,.
\end{equation}
In particular, these inequalities and Eqs.~\eqref{abcond} and
\eqref{abcond2} imply that $H^*$ cannot have both odd and even
parity algebraic eigenfunctions for any values of the parameters.
\end{remark}
Let us turn now to the algebraic spectrum of the $B_N$-type
Sutherland spin model \eqref{HBNSuth}. First of all, if
Eqs.~\eqref{abcond} and \eqref{abcond2} are satisfied the
wavefunctions in $\cM_{m_1}$ are normalizable and well behaved
near the singular hyperplanes. Secondly, if $\vp\in\cR_{m_1}$ is
an eigenfunction of $H$ with eigenvalue $E$ and $\ket s\in\cS$ is
an arbitrary spin state, it follows from Eq.~\eqref{HstH} that
$\La\big(\vp \ket s\big)\in\cM_{m_1}$ is either zero or an
eigenfunction of $H^*$ with the same energy $E$. The algebraic
spectrum of $H^*$ is thus a subset of the algebraic spectrum of
$H$. Instead of studying the conditions under which $\La\big(\vp
\ket s\big)$ does not vanish, in the next proposition we shall
directly construct from $\cB_{m_1}$ a basis of $\cM_{m_1}$ with
respect to which the matrix of $H^*|_{\cM_{m_1}}$ is upper
triangular. We shall use the following notation:
\begin{equation}
\label{m0} m_0 =
\overline{\vphantom{N^N}\frac{N-\overline{M}}{2M+1}}\,,
\end{equation}
where $\overline x$ denotes the smallest integer greater than or
equal to $x\in\RR$.
\begin{proposition}
\label{prop.basis} The $H^*$-invariant space $\cM_{m_1}$ is
nonzero if and only if $m_0\le m_1$. If this condition holds, a
basis of $\cM_{m_1}$ consists of states of the form
\begin{equation}\label{monbasis}
\La\big(f_n\ket{s_1,\dots,s_N}\big)\,,
\end{equation}
where $n\in\big[\ZZ^N\big]$ and $\ket{s_1,\dots,s_N}$ satisfy:
\begin{align}
i&)\quad \#(n_i)\le\begin{cases}
2M+1\,,& \text{if}\quad 0<n_i\le m_1\\[1mm]
\overline{M}\,,& \text{if}\quad n_i=0\,;
\end{cases}\label{cond1}\\[1mm]
ii&)\quad s_i>s_j\,,\quad \text{if}\quad
n_i=n_j\quad\text{and}\quad i<j\,;
\qquad\qquad\label{cond2}\\[1mm]
iii&)\quad s_i> 0\,,\quad \text{if}\quad n_i=0\,.\label{cond3}
\end{align}
\end{proposition}
\begin{proof}
In the first place, since $\La W = \ep(W)\La$ for any element $W$
of the realization of $\cW_N$ generated by $\Pi_{ij}$, $\Pi_i$,
the space $\cM_{m_1}$ is spanned by states of the form
$\La\big(f_n\ket s\big)$, where $n\in\big[\ZZ^N\big]$ and $\ket
s\in\cS$. Moreover, from the definition of the total
antisymmetriser $\La$ it follows that a state of the
form~\eqref{monbasis} with $n\in[\ZZ^N]$ vanishes if and only if
either $s_i=s_j$ when $n_i=n_j>0$ and $i\neq j$, or $s_i=\pm s_j$
when $n_i=n_j=0$ and $i\neq j$, or $s_i=0$ when $n_i=0$. In
particular, the condition~\eqref{cond1} is necessary to ensure
that the state~\eqref{monbasis} does not vanish. Since this
condition cannot hold if $n_1<m_0$, and $n_1\leq m_1$ for all
states in $\cM_{m_1}$, it follows that $\cM_{m_1}$ is trivial if
$m_1<m_0$.

On the other hand, if $m_0\leq m_1$ all the
states~\eqref{monbasis}--\eqref{cond3} are nonzero, and it is
immediate to show that they are also linearly independent.
Moreover, any nonzero state of the form~\eqref{monbasis} with
$n\in[\ZZ^N]$ can be written as
$$
\La\big[W\big(f_n\ket{s_1,\dots,s_N}\big)\big]=
\ep(W)\,\La\big(f_n\ket{s_1,\dots,s_N}\big)\,,
$$
where $n$ and $\ket{s_1,\dots,s_N}$
satisfy~\eqref{cond1}--\eqref{cond3}, and $W\in\cW_N$ is an
element of the stabilizer of $f_n$.
\end{proof}
\begin{corollary}
If $m_0\le m_1$, the dimension of the $H^*$-invariant space
$\cM_{m_1}$ is given by
\begin{equation}
\label{dim} \dim\big(\cM_{m_1}\big)=\binom{m_1(2M+1)+\overline
M}N\,.
\end{equation}
\end{corollary}
\begin{proof}
Indeed, from Eqs.~\eqref{cond2}--\eqref{cond3} it easily follows
that
\[
\dim\big(\cM_{m_1}\big)=\sum_{N_0+\dots+N_{m_1}=N}
\binom{\overline M}{N_0}\binom{2M+1}{N_1}\dots\binom{2M+1}{N_{m_1}}
=\binom{m_1(2M+1)+\overline M}N\,.
\]
\end{proof}
The algebraic spectrum of the hyperbolic $B_N$ Sutherland spin
model \eqref{HBNSuth} follows directly from Proposition
\ref{prop.basis}:
\begin{theorem}
\label{thm.spec} If $m_0\le m_1$, the algebraic energies of $H^*$
are given by
\begin{equation}
\label{phenergy} E_n^*=-\sum_i\la_{n,i}^2\,,
\end{equation}
where $n\in\big[\ZZ^N\big]$ satisfies the condition~\eqref{cond1}
and $\la_{n,i}$ is given by Eq.~\eqref{la}.
\end{theorem}
\begin{proof}
Let $\psi_{n,s}=\La\big(f_n\ket{s_1,\dots,s_N}\big)$ be an element
of the basis \eqref{monbasis}--\eqref{cond3} of $\cM_{m_1}$. Using
Eqs.~\eqref{Hgfn} and \eqref{HstH} we easily obtain
\begin{equation}
\label{Hstfn} H^* \psi_{n,s} = -\sum_i\la_{n,i}^2 \psi_{n,s} +
\sum_{\substack{n'\in \ZZ^N\\ n'\prec
\,n}}c_n^{n'}\,\La\big(f_{n'}\ket{s_1,\dots,s_N}\big)\,.
\end{equation}
The state $\La\big(f_{n'}\ket{s_1,\dots,s_N}\big)$ is proportional
to a basis element of the form $\psi_{n'',s'}$, with
$f_{n''}=Wf_{n'}$ for some element $W$ of $\cW_N$ and $n''\prec
n$. Therefore the matrix of $H^*|_{\cM_{m_1}}$ in the basis
\eqref{monbasis}--\eqref{cond3} is also upper triangular, with
diagonal elements given by \eqref{phenergy}.
\end{proof}
The algebraic ground state of the Hamiltonian $H^*$ and its
degeneracy $d$ can be determined using
Proposition~\ref{prop.basis} and the previous theorem:
\begin{proposition}
The multiindex $n$ yielding the algebraic ground state and the
degeneracy of this state are given by\\
i) \: $\ds N\le \overline M$:\qquad
$n=0\,,\quad d=\ds\binom{\overline M}N\,;$\\
ii) \: $\ds N>\overline M$:\vspace*{-5mm}
\begin{align}
& n = \big(\overbrace{\vphantom{1}m_0,\dots,m_0}^r,
\overbrace{m_0-1,\dots,m_0-1}^{2M+1}
,\dots,\overbrace{1,\dots,1}^{2M+1},
\overbrace{0,\dots,0}^{\overline M}\big),\label{n}\\[2mm]
& d= \binom{2M+1}r\,,\notag
\end{align}
\hphantom{ii) \: } where $r=N-(m_0-1)(2M+1)-\overline M$.
\end{proposition}
\begin{proof}
Note, first of all, that from the definition \eqref{m0} of $m_0$
it follows that $1\le r\le 2M+1$. Let $n\in\big[\ZZ^N\big]$ be a
nonnegative and nonincreasing multiindex satisfying $n_1\le m_1$
and Eq.~\eqref{cond1}. The contribution to the algebraic energy
\eqref{phenergy} of all the terms $\la_{n,i}^2$ such that $n_i$ is
equal to a fixed value $k\in\{n_j\}_{j=1}^N$ can be easily
evaluated in closed form. Indeed, denoting for brevity
\begin{gather*}
i_0=\ell(k)\,,\qquad i_1=\ell(k)+\#(k)-1\,,\\
\al = -\frac1{2a}\big(b+b'+2a(N-1)\big)\,,\qquad \al_k =
\al-\frac{k}a\,,
\end{gather*}
from \eqref{la} we have, for $k>0$,
\begin{align}
-\sum_{i=i_0}^{i_1} \la_{n,i}^2 &=
-4a^2\, \sum_{i=i_0}^{i_1}(i-i_0-i_1-\al_k+1)^2\notag\\
&= -4a^2\#(k)\Big[\al_k^2 + \big(i_0+i_1-2\big)\al_k
+\frac13\big(i_0^2+i_0i_1+i_1^2\big) -\frac16
\big(7i_0 + 5 i_1\big)+1\Big].\label{sumni}
\end{align}
Similarly, the contribution to the algebraic energy of all the
terms in Eq.~\eqref{phenergy} with $k=0$
\begin{equation}
-\sum_{i=i_0}^{i_1} \la_{n,i}^2 = -4a^2\,
\sum_{i=i_0}^{i_1}(i+\al-1)^2
\end{equation}
is easily seen to equal the right-hand side of Eq.~\eqref{sumni},
since $\al_0=\al$. The derivative of the right-hand side of
Eq.~\eqref{sumni} with respect to $k$, with $i_0$ and $i_1$ fixed,
is given by
\begin{equation}
4a\#(k)(2\al_k+i_0+i_1-2)\,.
\end{equation}
This is clearly positive, since $i_0+i_1-2\ge0$ and
$\al_k\ge\al-\frac {m_1}a>0$ on account of \eqref{m1}. Hence the
energy decreases if $k$ decreases, $i_0$ and $i_1$ being fixed. It
follows that any multiindex $n$ corresponding to the minimum value
of the algebraic energy must be of the form
\begin{equation}
\label{nstep} n =
\big(m,\dots,m,m-1,\dots,m-1,\dots,\ep,\dots,\ep\big)\,,
\end{equation}
where $\ep=0$ or $\ep=1$ and $m_0\le m\le m_1$.

Let $k$ be an integer in the range $1$ to $m$. We shall consider
next the change in the algebraic energy associated to the
multiindex \eqref{nstep} when $\#(k)$ decreases by $1$, while
$\#(k-1)$ increases by $1$ (including the case in which $k=\ep=1$
and therefore $\#(k-1)=\#(0)=0$). Suppose, for instance, that
$k\ge 2$. Denoting $i_2=\ell(k-1)+\#(k-1)-1$, the change in the
algebraic energy is given by
\begin{equation}
\begin{aligned}
&4a^2\bigg(-\sum_{i=i_0}^{i_1-1}(i-i_0-i_1-\al_k+2)^2
-\sum_{i=i_1}^{i_2}(i-i_1-i_2-\al_{k-1}+1)^2\\
&\qquad\:+\sum_{i=i_0}^{i_1}(i-i_0-i_1-\al_k+1)^2
+\sum_{i=i_1+1}^{i_2}(i-i_1-i_2-\al_{k-1})^2
\bigg)\\
&=-4\big(1+2a(\al_k+i_1-1)\big)\le -4(1+2a\al_k)<0\,.
\end{aligned}
\end{equation}
It may be similarly verified that when $k=1$ and either $\ep=0$ or
$\ep=1$ the change in the algebraic energy is negative. This
implies that the multiindex $n$ yielding the algebraic ground
state is of the form~\eqref{n} if $N>\overline M$, and zero
otherwise. The degeneracy of the algebraic ground state then
follows immediately from Proposition~\ref{prop.basis}.
\end{proof}
The algebraic ground energy can be easily obtained from the
previous proposition and Eqs.~\eqref{la} and~\eqref{energy}.
Indeed, denoting
$$
\nu=2M+1\,,\qquad c=b+b'+2m_0-a\,,
$$
the algebraic ground energy for even $\nu$ (that is, half-integer
$M$) is given by
$$
E^*_{0,e}=\frac13\,\nu m_0\big(4m_0^2(a\nu-1)-6cm_0-a\nu-2\big)
+\frac13\,(a^2-3c^2)N+2acN^2-\frac43\,a^2N^3\,,
$$
while for odd $\nu$ (integer $M$), the algebraic ground energy
reads
$$
E^*_{0,o}=E^*_{0,e}+m_0\big(a+2c+2m_0(1-a\nu)\big)\,.
$$

\section{The $B_N$-Type Sutherland Spin Chain}
\label{chain}
In this section we shall introduce a quantum spin chain closely
related to the hyperbolic $B_N$ Sutherland spin model
\eqref{HBNSuth}. We shall establish the integrability of this
chain by explicitly constructing a complete family of commuting
integrals of motion associated with the integrals $I_p^*$ of the
Hamiltonian \eqref{HBNSuth}.

The starting point in this construction is the following expansion
of the hyperbolic $B_N$ Sutherland spin Hamiltonian
\eqref{HBNSuth} in terms of the parameter $a$:
\begin{equation}\label{Hexp}
  H^*=-\sum_i\pa_{x_i}^2+a\,\cH^*+a^2\,U(\bx)\,,
\end{equation}
where
\begin{align}
\label{YamSC} &\cH^* =\sum_{i\ne j}\Big[\sinh^{-2} x_{ij}^-\,S_{ij}
+\sinh^{-2} x_{ij}^+\,\tS_{ij}\Big]
+\sum_i\big(\be\,\sinh^{-2} x_i-\be'\,\cosh^{-2} x_i\big)S_i\\
\label{U} &U(\bx)=\sum_{i\neq j}\big(\sinh^{-2} x_{ij}^- +
\sinh^{-2} x_{ij}^+\big)+ \sum_i\big(\be^2\sinh^{-2}\!x_i
-\be'^2\cosh^{-2}\!x_i\big)
\end{align}
and
$$
\be=\frac ba\,,\qquad \be'=\frac {b'}a\,.
$$
The Hamiltonian of the {\em hyperbolic Sutherland spin chain of
$B_N$ type} is by definition the operator $\szcH$, where the
superscript $0$ means that the coordinates $x_i$ are replaced by
the equilibrium points $x_i^0$ of the potential $U$, which satisfy
the system
\begin{equation}\label{eqpoints}
  \frac{\pa U}{\pa x_i}(x_1^0,\dots,x_N^0)=0\,,\qquad i=1,\dots,N\,.
\end{equation}
A necessary condition for the system \eqref{eqpoints} to have a
solution in the region $x_i>0$, $i=1,\dots, N$, is that
$\be'^2>\be^2+2(N-1)$. In fact, there is strong numerical evidence
that a solution exists if and only if $|\be'|>|\be|+2(N-1)$. Note
that this inequality corresponds to the condition~\eqref{abcond2}
(when $b>0$) or \eqref{abcond3} (when $b<0$) necessary for the
existence of square-integrable algebraic eigenfunctions of the
dynamical model, a fact certainly deserving further study.

Let us define the operator $\cJ_i\in C^\infty(\RR^N)\otimes\fK$ by
\begin{equation}
J_i = \pa_{x_i}-a\,\cJ_i\,,\qquad i=1,\dots, N\,.
\end{equation}
We shall also denote
\begin{equation}
\label{cIp} \cI_p=\sum_i\cJ_i^{2p},\qquad p\in\NN\,.
\end{equation}
Note that $\cI_1$ is the coefficient of $a^2$ in $I_1=-H$, and
thus equals $-U(\bx)$ by Eq.~\eqref{Hexp}. We shall prove below
that the operators $\{\cI^{*0}_p\}_{p=1}^N$ form a complete family
of commuting integrals of motion for the Sutherland $B_N$ spin
chain Hamiltonian $\szcH$. Let us begin by establishing the
commutativity of the operators $\cI_p^{*0}$ for all $p\in\NN$. In
fact, the following stronger result holds:
\begin{proposition}
The operators $\cI^*_p$ $(p\in\NN)$ form a commuting family.
\end{proposition}
\begin{proof}
The proposition follows directly from the commutativity of the
operators $I_p^*$, taking into account that $\cI_p^*$ is the
coefficient of $a^{2p}$ in the expansion of $I_p^*$ in powers
of~$a$.
\end{proof}

We show next that $\cI_p^{*0}$ commutes with $\szcH$ for all
$p\in\NN$. Note that this is not a consequence of the previous
proposition, since $\szcH$ is not proportional to
$\cI_1^{*0}=-U(\bx ^0)$.

\begin{proposition}
The operators $\cI_p^{*0}$ $(p\in\NN)$ commute with the $B_N$
Sutherland spin chain Hamiltonian $\szcH$.
\end{proposition}
\begin{proof}
From~\eqref{H} it follows that
$$
[H,J_i]=0\,,\qquad i=1,\dots,N\,.
$$
Using Eqs.~\eqref{Jexp} and~\eqref{Hexp} in the previous identity
and equating to zero the coefficient of $a^2$ in the resulting
expression we obtain
$$
[\cH\kern 1pt,\cJ_i]=-\frac{\pa U}{\pa x_i}\,,\qquad
i=1,\dots,N\,.
$$
It follows that
$$
[\cH\kern 1pt,\cI_p]=-\sum_{i=1}^N\sum_{r=0}^{2p-1} \binom{2p-1}r
\cJ_i^r\,\frac{\pa U}{\pa x_i}\,\cJ_i^{2p-r-1}\equiv \cC_p\,.
$$
The operator $\cC_1=-[\cH,U(\bx)]$ vanishes identically, since
$\cH$ does not contain partial derivatives and $U(\bx)$ is a
symmetric even function of $\bx$. Note, however, that $\cC_p$ need
not vanish for $p>1$. Expanding in powers of $a$ the identities
$[H,\La]=[I_p,\La]=0$ we obtain
$$
[\cH,\La]=[\cI_p,\La]=0\,,\qquad p\in\NN\,.
$$
By Lemma \ref{lem.comm} we have
\begin{equation}\label{commHIirr}
    \big[\scH,\scI_p\big]=\scC_p\,,\qquad p\in\NN\,.
\end{equation}
>From the symmetry of the function $U(\bx)$ with respect to
permutations and sign changes it follows that $\pa U/\pa x_i$
commutes with $K_{jk}$ and $K_j$ for $j,k\ne i$, while
\begin{align*}
K_{ij}\,\frac{\pa U}{\pa x_i} = \frac{\pa U}{\pa
x_j}\,K_{ij}\,,\qquad K_i\,\frac{\pa U}{\pa x_i} = -\frac{\pa
U}{\pa x_i}\,K_i\,.
\end{align*}
>From these identities one can easily show that $\scC_p$ is of the
form
$$
\scC_p=\sum_i \frac{\pa U}{\pa x_i}\,\cC_{p,i}^*\,,\qquad
p\in\NN\,.
$$
The commutativity of $\szcH$ with $\cI_p^{*0}$ follows from the
latter equation, Eq.~\eqref{commHIirr} and the definition of the
equilibrium points \eqref{eqpoints}.
\end{proof}
Finally, we prove the algebraic independence of the set
$\{\cI^{*0}_p\}_{p=1}^N$, thus establishing the integrability of
the hyperbolic Sutherland spin chain of $B_N$ type:
\begin{theorem}
The operators $\{\cI_p^{*0}\}_{p=1}^N$ form a complete family of
commuting integrals of motion for the $B_N$ Sutherland spin chain
Hamiltonian $\szcH$.
\end{theorem}
\begin{proof}
The set $\{\cI_p^0\}_{p=1}^N$ is algebraically independent, since
the operators $\cJ_i^0$ ($i=1,\dots,N$) are linearly independent
and commute with each other. The counterpart of Lemma
\ref{lem.fund} for the inverse of the star operator implies that
the family $\{\cI_p^{0*}\}_{p=1}^N$ is also algebraically
independent. The lemma now follows from the identity $\cI_p^{0*}
=\cI_p^{*0}$.
\end{proof}

Note also that the constants of motion $\cI_p^{*0}$ ($p\in\NN$)
commute with the total permutation and sign reversing operators
$\Pi_{ij}$ and $\Pi_i$. This follows from the identities
$[I_p^*,\Pi_{ij}]=[I_p^*,\Pi_i]=0$ by taking the coefficient of
$a^{2p}$.

\begin{acknowledgments}
This work was partially supported by the DGES under grant
PB98-0821. R. Zhdanov would like to acknowledge the financial
support of the Spanish Ministry of Education and Culture during
his stay at the Universidad Complutense de Madrid. The authors
would also like to thank the referee for several helpful remarks.
\end{acknowledgments}

\end{document}